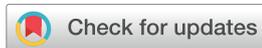

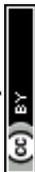

# New insights into the 1D carbon chain through the RPA

Benjamin Ramberger and Georg Kresse*

We investigated the electronic and structural properties of the infinite linear carbon chain (carbyne) using density functional theory (DFT) and the random phase approximation (RPA) to the correlation energy. The studies are performed *in vacuo* and for carbyne inside a carbon nano tube (CNT). In the vacuum, semi-local DFT and RPA predict bond length alternations of about 0.04 Å and 0.13 Å, respectively. The frequency of the highest optical mode at the $\Gamma$ point is 1219 cm$^{-1}$ and about 2000 cm$^{-1}$ for DFT and the RPA. Agreement of the RPA to previous high level quantum chemistry and diffusion Monte-Carlo results is excellent. For the RPA we calculate the phonon-dispersion in the full Brillouine zone and find marked quantitative differences to DFT calculations not only at the $\Gamma$ point but also throughout the entire Brillouine zone. To model carbyne inside a carbon nanotube, we considered a (10,0) CNT. Here the DFT calculations are even qualitatively sensitive to the $k$-points sampling. At the limes of a very dense $k$-points sampling, semi-local DFT predicts no bond length alternation (BLA), whereas in the RPA a sizeable BLA of 0.09 Å prevails. The reduced BLA leads to a significant red shift of the vibrational frequencies of about 350 cm$^{-1}$, so that they are in good agreement with experimental estimates. Overall, the good agreement between the RPA and previously reported results from correlated wavefunction methods and experimental Raman data suggests that the RPA provides reliable results at moderate computational costs. It hence presents a useful addition to the repertoire of correlated wavefunction methods and its accuracy clearly prevails for low dimensional systems, where semi-local density functionals struggle to yield even qualitatively correct results.

## 1 Introduction

Carbyne – the infinite sp$^1$ hybridized carbon chain – is a model 1D material with fascinating properties that point to promising applications. For example, it is anticipated that new composite materials may be able to exploit the unique nature of carbyne and provide novel mechanical functionalities.[1] Another exciting application might be realized by using single chains of linear carbon as a channel for field-effect transistors, thus bringing the channel width to the extreme minimum of one carbon atom thickness.[2] Unfortunately, carbyne is extremely unstable and very reactive at ambient conditions, making its synthesis and detection experimentally very difficult. While for many years, the use of heavy end-capping groups for stabilization was very common to synthesize linear carbon chains, the lengths that could be achieved were very limited, *e.g.* 44 atoms in a 2010 study by Chalifoux and Tykwinski.[3] Another approach is to use carbon nanotubes (CNTs) as confining nanoreactors to synthesize 1D materials. In a 2016 study, L. Shi *et al.* reported a method for the bulk production of long acetylenic linear carbon chains protected by thin double-walled CNTs. They showed that stable chains composed of more than 6000 carbon atoms can be produced.[4] This important step towards the bulk production of carbyne is not only encouraging from an experimental perspective, but also shifts the focus of theoretical investigations towards linear carbon chains confined by CNTs.

Theoretical studies on the ideal carbon chain in vacuum show that Peierls distortions[5] give rise to bond length alternations (BLA) that are associated with single and triple C–C bonds. The underlying reason is that the carbon chain is characterized by a half-filled band at the Fermi level, and a small distortion from the equidistant chain introduces a gap near the Fermi level, thus lowering the total energy of the electrons. Although conventional local and semi-local density functionals reproduce this effect qualitatively, there is a considerable quantitative discrepancy for the predicted BLAs between correlated wave function methods like coupled cluster singles and doubles with perturbative triples [CCSD(T)] or diffusion Monte-Carlo (DMC) and (semi) local functionals, such as LDA or PBE. Hybrid functionals, for instance HSE06, provide an improvement over the (semi) local functionals, but the carbyne *in vacuo* BLA from DMC is still around 50% larger than for HSE06.[6] Likewise, the phonon frequencies are not accurately predicted by (semi) local functionals.[6,7]

*University of Vienna, Faculty of Physics and Center for Computational Materials Sciences, Kolingasse 14-16, 1090 Vienna, Austria. E-mail: georg.kresse@univie.ac.at*

These findings already indicate that, although carbyne is a prototypical seemingly simple 1D model system, one needs to go beyond standard semi-local DFT to quantitatively describe the phenomena exhibited by it.

For the theoretical study of carbyne inside CNTs, which is of particular interest due to the before mentioned pathway for the production of bulk carbyne, the situation is even more involved, since van-der-Waals (vdW) interactions between the CNTs and the linear carbon chain come into play. Wanko *et al.* reported that two distinct and separable effects lead to a red shift of 118–290 cm$^{-1}$ in the Raman spectra of carbon chains inside CNTs compared to *in vacuo* frequencies. On the one hand, a charge transfer (CT) from the tube to the chain softens the chains Γ mode. On the other hand, vdW interactions lead to a similar softening that potentially even exceeds the one from the CT.[7] Since vdW interactions are not accounted for by (semi) local density functionals, the carbyne–CNT system clearly cannot be described accurately by semi-local functionals.

A promising method to tackle this materials problem is provided by the RPA. It has been shown that the RPA performs very well for different bonding types, including vdW interactions.[8–10] The availability of low-scaling $\mathcal{O}(N^3)$ RPA algorithms[11,12] as well as the possibility to calculate forces in the RPA[13] offers a potential route to accurately describe the carbyne–CNT system at moderate computational cost. Additionally, versatile tools for the RPA that would facilitate further investigations have been reported recently. On the one hand, RPA natural orbitals can be employed for the efficient calculation of accurate second-order Møller Plesset (MP2) or full configuration interaction (FCI) energies.[14] On the other hand, using the RPA to benchmark density functionals[15] might lead to the identification of a sufficiently accurate functional that can be used for large scale structure studies. Finally, the very recent introduction of a finite-temperature generalization of the low-scaling RPA algorithms[16] provides an elegant way to treat the partial occupancies that occur due to the charge transfer from the tube to the chain. These prospects lead us to believe that investigating the performance of the RPA for the carbyne–CNT system will yield a valuable contribution to further our knowledge on this intriguing material.

## 2 Computational details

We calculated PBE[17] and RPA[11,12,16] total energies for different BLAs of carbyne *in vacuo* as well as inside a (10,0) (CNT). Additionally we checked how adding vdW-corrections (PBE+D3), as proposed by Grimme *et al.*,[18] affects the energy surface. Throughout this work we used VASP 6.1[19,20] with its implementation of the projector augmented wave (PAW) formalism,[21,22] a 400 eV plane wave cut-off, and the potential C_s_GW. This potential uses three projectors for the 2s and 2p partial waves and a truncated all-electron potential as local potential. The radial cutoffs are 1.35 a.u., 1.75 a.u and 0.7 a.u. for the s, p, and local potential, respectively.

### 2.1 Geometries and *k*-points

As the "standard" cell, we employed an 11.200 Å × 11.200 Å × 12.875 Å tetragonal cell with periodic boundary conditions in which the carbyne's and CNT's axes were aligned along the *z*-axis. The choice of this cell resulted from preliminary calculations: the geometry of the carbyne chain was determined by varying the lattice constant in the periodic direction between 2.5 and 2.7 Å, and relaxing the internal coordinate (BLA) for each lattice constant. For the equilibrium lattice constant, a value of 2.575 Å and 2.590 Å was determined with PBE and RPA, respectively.

The experimental bond length of carbon in graphene is about 1.42 Å. For the (10,0) zig-zag tube, this yields a lattice constant of 4.26 Å. If the tube is three times replicated this yields a lattice constant of 12.780 Å. Thus, a 5-fold super cell of carbyne fits almost perfectly into the three times replicated tube. Since we wanted to keep the strain on the carbyne chain and the tube at a low level, we decided to fix the lattice constant to 12.875 Å. This corresponds to the five-fold supercell of carbyne in PBE, and hence the carbyne is strain free at the level of PBE. This lattice constant is also very close to the arithmetic mean of the experimental estimates for the three times replicated (10,0) tube and the five-fold replicated primitive carbyne chain in the RPA (12.95 Å). All the other coordinates were relaxed at the PBE level until the forces on all atoms were smaller than $10^{-5}$ eV Å$^{-1}$.

In total there are 130 C-atoms in the filled standard cell-120 CNT C-atoms around 10 carbyne C-atoms, which are placed on the CNT's principal axis (Fig. 1).

For the carbyne *in vacuo* calculations, we used a primitive 11.2 Å × 11.2 Å × 2.575 Å tetragonal cell with two atoms at (0, 0, BLA/2) Å and (0, 0, 1.2875) Å. With this setup, the number of *k*-points in *z*-direction needs to be five times as great as in the standard cell to allow a one-to-one comparison.

In the following, when we specify the *k*-point sampling, we will always refer to the number of *k*-points per primitive cell – kpc. For example, 2 *k*-points in the standard cell correspond to 10 kpc.

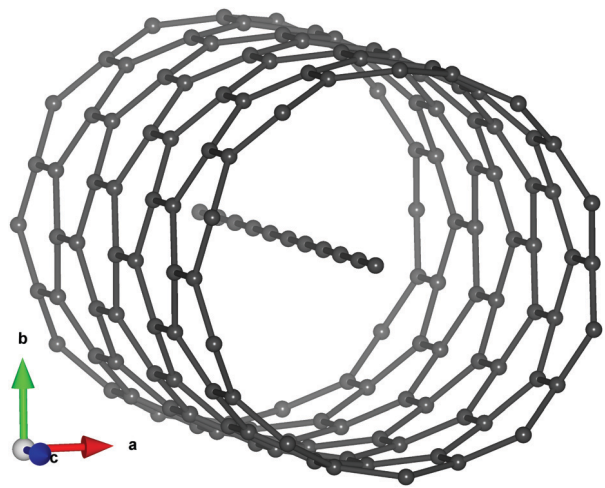

Fig. 1 Standard cell used in this work: carbyne in (10,0) CNT. Graphic created using VESTA.[23]

We used different $\Gamma$-centered $1 \times 1 \times k_z$ $k$-meshes with $k_z$ ranging from 10 to 640 kpc for PBE and from 10 to 80 (40) kpc for the RPA *in vacuo* (in the CNT).

## 2.2 RPA

In general, we performed the RPA calculations on top of PBE, *i.e.* we use occupied and virtual PBE orbitals to calculate the correlation energy perturbatively in the RPA.[24] For the RPA calculations of carbyne *in vacuo* the standard low scaling $\mathcal{O}(N^3)$ implementation in VASP[11,12] was used with 12 imaginary time/frequency grid points. However, for the filled standard cell, we used a FTRPA approach[16] that is particularly useful for systems with partial occupancies (in the one-electron orbitals of the underlying mean-field ground state). In this method, the ground state is calculated with Fermi smearing, *i.e.* the occupancies follow a Fermi distribution that is controlled *via* the smearing parameter $\sigma = k_B T[\text{eV}]$. The RPA correlation is then calculated on an optimized imaginary time/frequency grid for the given $\sigma$. In this work, we calculated and compared BLA *vs.* FTRPA energy curves of carbyne inside a (10,0) CNT for $\sigma$ [eV] $\in$ {0.03,0.05,0.10}, using 16 time/frequency grid points.

## 2.3 Phonon frequencies

The full phonon dispersion was calculated for carbyne *in vacuo* using DFT and the RPA, finite displacements and deriving from the forces the inter-atomic force constants.[13] For PBE, the primitive cell was replicated 20 times, whereas for RPA a more modest replication of 5 unit cells was chosen. To our knowledge, this is the first time that the phonon dispersion relation of carbyne was calculated using a correlated wavefunction method.

For carbyne inside the nanotube, calculating the phonon dispersion would be exceedingly demanding, since the translational symmetry is broken. Therefore, we only determined the highest frequency mode using a finite displacement corresponding to the longitudinal optical (LO) mode. To estimate $\omega_\Gamma$, we used the harmonic approximation for the energy $E(b)$ as a function of the BLA $b$

$$E(b) = E_0 + \frac{1}{2}\mu_C \omega_\Gamma^2 \left(\frac{b}{2} - \frac{b_0}{2}\right), \quad (1)$$

with the reduced mass of two carbon atoms $\mu_C = 6\ \mu$, and the BLA (energy) at equilibrium being $b_0$ ($E_0$). We used a polynomial fit for the total energy $f_E(b)$ and calculated the equilibrium BLA as

$$\left.\frac{\partial f_E(b)}{\partial b}\right|_{b_0} = 0, \quad (2)$$

and the phonon frequency $\omega_\Gamma$ as

$$\omega_\Gamma = \left(\frac{4}{\mu_C}\left.\frac{\partial^2 f_E(b)}{\partial b^2}\right|_{b_0}\right)^{1/2}. \quad (3)$$

For all PBE energy curves and for the *in vacuo* RPA energy curves sixth-order polynomials were used, since the data were very smooth. For the less smooth FTRPA energy curves, we had to resort to third-order polynomial fits.

# 3 Results and discussion

## 3.1 Preliminary tests

As mentioned above, the choice of our filled standard cell resulted from the ratio of the primitive vectors in $z$-direction, $a_z$, for the relaxed chain in vacuum and the empty (10,0) CNT. Since the ratio is very close to 3:5, our choice of the filled standard cell provided an ideal starting point for the investigation of the carbyne–CNT system under periodic boundary conditions. To start with, we confirmed, on the PBE level, that our choice of the standard cell minimizes the total energy of the filled standard cell with respect to the cell parameters. Additionally we studied how the total energy of the carbyne–CNT system changes, when the carbon chain as a whole is shifted along the principal axis. The differences were below 1 meV and are thus small but not entirely negligible. Since we considered the geometry of the tube to be fixed in its equilibrium state, the chain's BLA is the only structure parameter left that significantly affects the energy of the system. Furthermore, we checked whether the primitive cell and the standard cell yield numerically identical results. *In vacuo*, the energy differences between a 5-fold super cell with $N$ $k$-points and a primitive cell with 5$N$ $k$-points and the same BLA and $x$–$y$-vacuum distance was smaller than 0.2 meV. Also phonon frequencies at the $\Gamma$ point are identical. Therefore, as expected, it is safe to compare results from the primitive cell of carbyne *in vacuo* with those of carbyne in the (10,0) CNT, as long as one uses the same kpc.

Using PBE+D3 vdW corrections, we found that with 640 kpc *in vacuo* and in our standard (10,0) CNT cell, the equilibrium BLA and LO $\Gamma$ frequencies changed by less than 0.2% compared to plain PBE. Therefore we did not pursue further investigations into vdW corrections, but report these results in Table 1.

## 3.2 BLA and phonons *in vacuo*

VASP uses an internal cut-off extrapolation to calculate the RPA energy. Here, we compare the BLAs and LO $\Gamma$ frequencies obtained by using the RPA energies at the highest cut-off for the response function (here 266.7 eV) with those obtained using the cut-off extrapolated energies (see Fig. 2). We found that the cut-off extrapolation slightly increases the BLA and the energy gain related to the BLA. This also brings the RPA results in better agreement with results from other correlated wavefunction methods such as DMC.[6] The difference between the highest cut-off and cut-off extrapolation in the predicted BLA and the LO $\Gamma$ frequency was, however, less than 5% and 1%, respectively, for carbyne *in vacuo*.

Table 1 BLAs and LO $\Gamma$ frequencies for carbyne. Note that the results in this table were obtained with different super cells and/or $k$-meshes. The kpc column indicates the equivalent number of $k$-points in the primitive cell

| Method | kpc | BLA [Å] | | $\omega_\Gamma$ [cm$^{-1}$] | |
| --- | --- | --- | --- | --- | --- |
| | | *In vacuo* | In CNT | *In vacuo* | In CNT |
| PBE | 640 | 0.037 | 0.000 | 1219 | 1112 |
| PBE+D3 | 640 | 0.037 | 0.000 | 1221 | 1112 |
| RPA | 80 | 0.129 | — | 2000 | — |
| (FT)RPA | 40 | 0.126 | 0.091 | 1960 | 1614 |
| DMC[6] | 16 | 0.136(2) | — | 2084(5) | — |
| CCSD(T)[7] | —[a] | 0.125 | | 2075 | — |
| Experiment[7] | — | — | — | — | 1800[b] |

[a] Obtained by extrapolation from finite molecules of up to 32 atoms.
[b] Obtained from long linear carbon chains inside double walled CNTs.

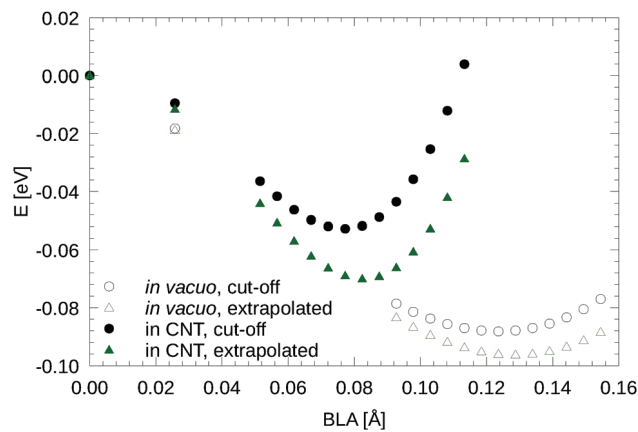

Fig. 2 Cut-off extrapolated RPA energy *vs.* BLA curves in comparison to the same curves without cut-off extrapolation.

In Fig. 3 on the left hand side, we show the BLA as well as the phonon frequencies for the LO $\Gamma$ frequency *in vacuo*. The results are very sensitive to the density of the *k*-point mesh, especially for the DFT (PBE) calculations. For example, for carbyne *in vacuo*, 32 and 64 kpc yield BLAs 27% respectively 5% larger than at 640 kpc, see Fig. 3(a). Likewise, the LO frequency at the $\Gamma$ point is strongly *k*-point dependent, see Fig. 3(c). Typically about 120 kpc are required to obtain highly accurate results. Remarkably the RPA results are somewhat less sensitive to the *k*-point sampling. The differences between 80 kpc and 40 kpc are quite small, and although we have attempted to use 80 kpc wherever possible, for the carbyne in the nanotube we had to reduce the sampling to 40 kpc (see below).

*In vacuo* using PBE, our predicted BLA is just 0.037 Å (with 640 kpc), while the RPA yields 0.129 Å (with 80 kpc), see Fig. 3(a). The RPA result is remarkably close to the 0.136(2) Å from DMC.[6] Considering that the *k*-point sampling was rather coarse in the DMC (equivalent to 16 kpc), the agreement is very satisfactory.

In Fig. 4 we show the full phonon dispersion relation for carbyne *in vacuo*, for the RPA and PBE. To be consistent with our other calculations, we have chosen a five-fold super cell with a lattice constant of 5 × 2.575 Å for the RPA calculations, whereas the PBE calculations were performed for a 20 times repeated unit cell. To confirm the results, in particular at the Brillouin zone boundary, we performed an additional RPA calculation (although with less *k*-points) in a four-fold super cell at the RPA lattice constant 4 × 2.590 Å (squares in Fig. 4).

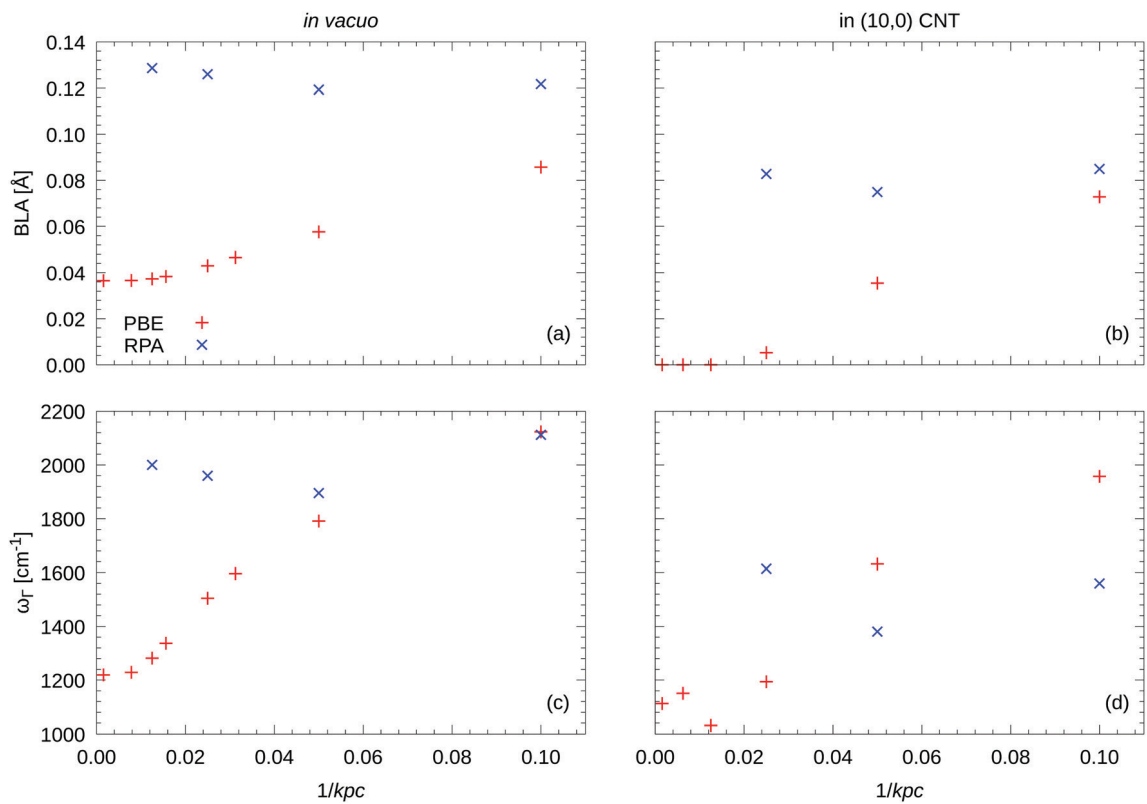

Fig. 3 *k*-Point convergence of equilibrium BLA and LO $\Gamma$ frequency $\omega_\Gamma$ for carbyne *in vacuo* and in CNT for PBE and (FT)RPA. RPA results *in vacuo* (panel a and c) were obtained with 12 time/frequency grid points. FTRPA results for carbyne in the (10,0) CNT (panel b and d) were obtained with a smearing parameter $\sigma$ = 0.1 eV and 16 time/frequency grid points.

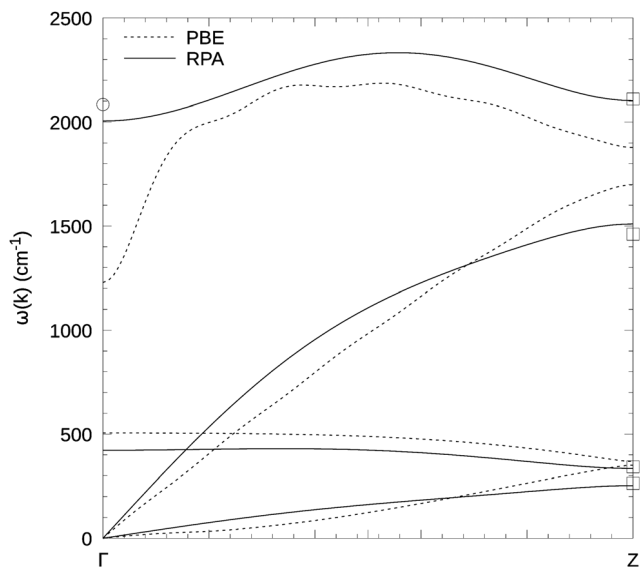

Fig. 4 Phonon dispersion relation for carbyne in vacuum. The RPA calculations (solid line) were performed in a 5-fold super cell with lattice constant 5 × 2.575 Å, 11.20 Å vacuum and 8 $k$-points (40 kpc). For PBE (dashed line) we used a 20-fold super cell with 8 $k$-points (160 kpc). Note that for PBE, we obtained the lowest acoustic mode with 4 $k$-points (80 kpc) in the 20-fold cell, because in the 160 kpc setup it exhibited imaginary frequencies. We added the DMC result (circle) at $\Gamma$ (2084 cm$^{-1}$)[6] and additional RPA results (squares) from a 4-fold super cell and 8 $k$-points (32 kpc) at $Z$.

It appears that the slight variation in the lattice constant has only a very small effect on the phonon dispersion. Likewise this test calculation demonstrates that the RPA force constants decay sufficiently fast that a five times replicated unit cell yields reliable results (a five times replicated unit cell does not guarantee technically accurate vibrational frequencies at the $Z$ point, since the corresponding phonon pattern is not commensurate with the supercell).

Both, Fig. 3 and 4, show that the LO $\Gamma$ frequency of carbyne *in vacuo* is very small at the level of PBE (1219 cm$^{-1}$). Compared to the DMC result of 2084(5) cm$^{-1}$ the error is almost 40%. As was the case for the BLA, the RPA is again quite close to DMC and wave function based results with 2000 cm$^{-1}$ at 80 kpc (DMC 2084 cm$^{-1}$,[6] extrapolated CCSD(T) around 2075 cm$^{-1}$ [7]). Again, it needs to be emphasized that the DMC results were obtained using supercells corresponding to 16 kpc, and the wave function based results are extrapolated from finite molecules with 32 atoms. At such low $k$-point densities, we also obtain somewhat larger phonon frequencies using the RPA.

### 3.3 BLA and phonons in (10,0) nanotube

Our PBE calculations for the filled standard cell confirm previous results.[25,26] The BLA vanishes for carbon chains inside single walled CNTs (SWCNT) when using (semi) local density functionals. However, we want to emphasize that this is again only true for sufficiently dense $k$-point grids (here $\geq 80$ kpc, compare Fig. 3). Likewise, the LO frequency at $\Gamma$ is strongly $k$-point dependent, see Fig. 3(d).

The suppression of the BLA in PBE is fairly easy to understand by inspecting the band structures in Fig. 5. For the non-dimerized carbyne chain in the nanotube, the Fermi level of the combined system is well above the Fermi-level of the carbon chain in the vacuum (cross in Fig. 5 left panel). This causes a significant charge transfer from the nanotube to the carbyne chain. Since the energy gain by the Peierls distortion can only be realized, if the upwards moving band is unoccupied and the downwards moving band is occupied, a charge transfer to the chain effectively suppresses the Peierls distortion. Also, recall that the Peierls distortion was small to begin with in DFT.

In contrast to the PBE calculations, we found that the carbyne's BLA does not vanish inside the tube using the RPA. This is a remarkable qualitative difference. Concerning the technical accuracy of this result, we have a few comments to make. As can be seen in Fig. 3(a), the equilibrium BLA for the primitive cell at 40 kpc is only slightly lower (2%) than using 80 kpc in vacuum. We are therefore quite confident that the reported BLAs obtained in the (10,0) CNT are reasonably accurate for 40 kpc. Depending on $\sigma$, the BLA of carbyne inside a (10,0) CNT calculated using the FTRPA is 0.083, 0.089 or 0.101 [Å] for $\sigma$ = 0.10, 0.05 or 0.03 [eV]. These broadenings correspond to temperatures of about 1200 K, 600 K and 400 K, hence a small width is usually preferable in order to compare with experimental results measured around room temperature. However in this case, the BLAs and LO $\Gamma$ frequencies do not change qualitatively, but the energy surface becomes smoother with increasing $\sigma$. The BLA is not particularly sensitive to the roughness of the energy surface, so that we consider 0.091 Å – the mean over different $\sigma$ – to be the best estimate.

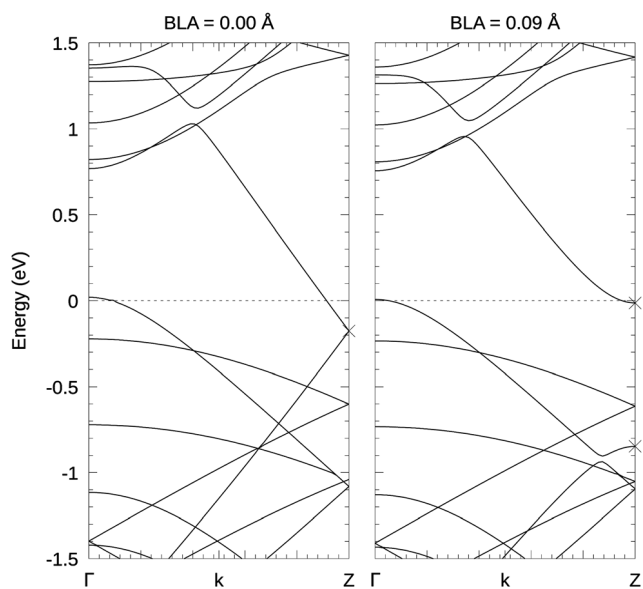

Fig. 5 PBE band structure of carbyne inside a (10,0) CNT. The left panel shows the band structure at vanishing BLA, which corresponds to the PBE equilibrium. The right panel shows the band structure at a BLA of 0.09 Å which corresponds to the average BLA for FTRPA. The bands from the carbyne chain are marked with crosses at the $Z$. Increasing charge transfer is correlated with decreasing BLA.

The band structure shown in Fig. 5 evaluated at the BLA of 0.09 Å – close to the RPA BLA – shows that the valence band edge of the nanotube and the conduction band edge of the carbyne are almost iso-energetic. This means that the charge transfer is small at the equilibrium RPA BLA.

However, this onset of charge transfer also makes it somewhat difficult to make accurate predictions for the carbyne's LO $\Gamma$ frequency. As the BLA becomes smaller, charge is progressively transferred from the tube to the carbyne chain. If the $k$-point density is not very large, the one electron occupancies can change somewhat discontinuously. Thus, the RPA calculations would certainly have benefited from denser $k$-point grids, however, more than 40 kpc for the FTRPA calculations of the filled standard cell would have been difficult with the available computational resources due to excessive memory requirements.

The LO $\Gamma$ frequency is rather sensitive with respect to the roughness of the energy surface and therefore, with $\sigma$ = 0.10 eV we obtained the most reliable polynomial fit and our best estimate for the LO $\Gamma$ frequency at 1614 cm$^{-1}$. Hence, the tube induced redshift of the carbyne's LO $\Gamma$ frequency is 346 cm$^{-1}$ at 40 kpc. This value is quite close to the 290 cm$^{-1}$ reported by Wanko et al.,[7] which was obtained by calculating the difference between the theoretical carbyne in vacuo frequency from extrapolated CCSD(T) and the experimentally observed frequencies of long linear carbon chains inside double-wall CNTs (DWCNTs). Since the PBE functional does not yield any BLA inside the tube, the energy surface becomes very soft, and the phonon frequencies are even further red-shifted to about 1100 cm$^{-1}$ using 640 kpc. Wanko et al.[7] related this erroneous description to the lack of vdW interactions in the PBE functional, but we do not agree with this interpretation. Semi-local functionals are not accurate for 1D systems, in particular, in the prediction of Peierls instabilities. Because of self-interaction errors, they generally underestimate the magnitude of the BLA, both in vacuo as well as in the tube. In vacuum, the disagreement is only quantitative, but since the charge transfer from the tube to the chain decreases the tendency for a bond length alteration (see Fig. 5), PBE ultimately fails to yield any BLA inside the tube, whereas in the RPA the BLA is only decreased but does not vanish entirely.

In agreement with the arguments given in the previous paragraph, vdW corrections do not change the BLA or vibrational frequencies. Table 1 summarizes our findings and compares them with the literature.

A final comment on the technical challenges of the present calculations seems in place. We performed our calculations on the Vienna Scientific Cluster (VSC4) using Intel Skylake Platinum 8174 cores. A 80 kpc carbyne in vacuo RPA energy calculation took typically 45 min with 144 cores in parallel and used up to about 5.1 GB RAM per core. For the filled standard cell and 20 kpc we used 192 cores in parallel that took typically 1 h 15 min for one FTRPA energy calculation and used up 4.6 GB RAM per core. With 40 kpc the memory demand increased to about 8 GB RAM, and 80 kpc calculations were not possible using 192 cores. The RPA phonon calculation for the five-fold primitive cell of carbyne in vacuo with 40 kpc were most expensive and took 36 h on 384 cores and used 6.9 GB RAM per processor. All in all, these are very reasonable computational requirements, except for the fairly excessive memory demands of the low scaling RPA code.

## 4 Conclusion

We have investigated the performance of the (FT)RPA for carbyne in vacuo and inside a SWCNT and found good agreement with experiments and other correlated wave function methods at moderate computational cost. Crucially, we found that a dense $k$-point sampling is absolutely required to obtain reliable and accurate results. Specifically, the PBE functional requires more than 128 $k$-points in the primitive cell to yield converged results. For example, for the carbyne chain inside the SWCNT, more than 80 kpc are necessary in order to observe that the BLA vanishes for PBE. The RPA does not show such a strong $k$-point dependence and reasonable results can already be expected around 40 kpc. This allowed us to calculate the full phonon dispersion relation of carbyne in vacuo. Remarkably, the results are quantitatively quite different from the PBE results, not only for the LO mode close to $\Gamma$ point but throughout the entire Brillouine zone. The important point is that previous calculations for diamond and graphene showed only very small differences between semi-local functionals and the RPA.[13] The increased discrepancy in 1D is a strong indication that for low dimensional systems, beyond mean field methods are required to predict highly accurate results.

Our most interesting result is however the predicted BLA for carbyne inside a carbon nanotube. Using the RPA, our best estimate for the BLA is 0.09 Å in a (10,0) CNT. Previous assessments were mostly indirect and relied on establishing correlations between empty and filled tubes and the predicted and observed LO $\Gamma$ frequency. The BLA predicted by the RPA seems to be in good agreement with the previous deductions, specifically the RPA predicts a LO $\Gamma$ frequency of 1614 cm$^{-1}$ in very reasonable agreement with the available experimental data (1800 cm$^{-1}$). As to why our LO $\Gamma$ frequency is too small, we can only speculate. First, the RPA might somewhat underestimate the phonon frequencies of the free carbyne chain (RPA 2000 cm$^{-1}$ other correlated wave functions methods around 2080 cm$^{-1}$, albeit using only few $k$-points or rather short finite carbyne fragments). The origin for this small difference could be related to the approximations that the RPA makes, for instance, the neglect of second-order exchange or errors in the PBE density matrix. An alternative explanation is that the energy surface is very anharmonic at small BLAs, so that one might need to account for anharmonic effects in the theoretical calculations or beyond adiabatic coupling of phonons to electrons at the Fermi-level. Finally, the amount of charge transferred to the carbyne chain determines how strongly the BLA is suppressed and how much the LO phonon frequency changes compared to the free standing chain. The charge transfer might be different in double wall tubes, as the one used in experiments, than in our simplified theoretical model.

On the upside, the present theoretical study clearly demonstrates the capabilities of the RPA and the recently presented FTRPA method.[16] Calculations for carbyne inside a nano-tube are simply not possible with any other correlated wave function method or diffusion Monte-Carlo. Using the RPA forces,[13] we were also able to predict the phonon dispersion relation of free standing carbyne *in vacuo* from beyond mean field methods, again, something hardly possible for other correlated wave function methods. Due to the lower symmetry inside the tube, calculations of the phonon dispersion relation for carbyne inside a SWCNTs are not yet possible, but might become possible in the not too distant future. We hope that our current results inspire more studies of 1D systems using the RPA and that the RPA can help to unravel the fascinating behavior of 1D materials.

## Conflicts of interest

There are no conflicts to declare.

## Acknowledgements

Computations were performed on the Vienna Scientific Cluster VSC4.

## References

1 M. Liu, V. I. Artyukhov, H. Lee, F. Xu and B. I. Yakobson, *ACS Nano*, 2013, **7**, 10075–10082.
2 R. H. Baughman, *Science*, 2006, **312**, 1009–1010.
3 W. A. Chalifoux and R. R. Tykwinski, *Nat. Chem.*, 2010, **2**, 967–971.
4 L. Shi, P. Rohringer, K. Suenaga, Y. Niimi, J. Kotakoski, J. C. Meyer, H. Peterlik, M. Wanko, S. Cahangirov, A. Rubio, Z. J. Lapin, L. Novotny, P. Ayala and T. Pichler, *Nat. Mater.*, 2016, **15**, 634–639.
5 R. Peierls, *Quantum theory of solids*, Oxford University Press, Oxford, 1955, pp. 108–112.
6 E. Mostaani, B. Monserrat, N. D. Drummond and C. J. Lambert, *Phys. Chem. Chem. Phys.*, 2016, **18**, 14810–14821.
7 M. Wanko, S. Cahangirov, L. Shi, P. Rohringer, Z. J. Lapin, L. Novotny, P. Ayala, T. Pichler and A. Rubio, *Phys. Rev. B: Condens. Matter Mater. Phys.*, 2016, **94**, 195422.
8 Y. S. Al-Hamdani, M. Rossi, D. Alfè, T. Tsatsoulis, B. Ramberger, J. G. Brandenburg, A. Zen, G. Kresse, A. Grüneis, A. Tkatchenko and A. Michaelides, *J. Chem. Phys.*, 2017, **147**, 044710.
9 J. A. Garrido Torres, B. Ramberger, H. A. Früchtl, R. Schaub and G. Kresse, *Phys. Rev. Mater.*, 2017, **1**, 060803.
10 J. G. Brandenburg, A. Zen, M. Fitzner, B. Ramberger, G. Kresse, T. Tsatsoulis, A. Grüneis, A. Michaelides and D. Alfè, *J. Phys. Chem. Lett.*, 2019, **10**, 358–368.
11 M. Kaltak, J. Klimes and G. Kresse, *J. Chem. Theory Comput.*, 2014, **10**, 2498–2507.
12 M. Kaltak, J. Klimes and G. Kresse, *Phys. Rev. B: Condens. Matter Mater. Phys.*, 2014, **90**, 054115.
13 B. Ramberger, T. Schäfer and G. Kresse, *Phys. Rev. Lett.*, 2017, **118**, 1–5.
14 B. Ramberger, Z. Sukurma, T. Schäfer and G. Kresse, *J. Chem. Phys.*, 2019, **151**, 214106.
15 M. Bokdam, J. Lahnsteiner, B. Ramberger, T. Schäfer and G. Kresse, *Phys. Rev. Lett.*, 2017, **119**, 1–5.
16 M. Kaltak and G. Kresse, *Phys. Rev. B*, 2020, **101**, 205145.
17 J. P. Perdew, K. Burke and M. Ernzerhof, *Phys. Rev. Lett.*, 1996, **77**, 3865–3868.
18 S. Grimme, J. Antony, S. Ehrlich and H. Krieg, *J. Chem. Phys.*, 2010, **132**, 154104.
19 G. Kresse and J. Hafner, *Phys. Rev. B: Condens. Matter Mater. Phys.*, 1993, **47**, 558–561.
20 G. Kresse and J. Furthmüller, *Phys. Rev. B: Condens. Matter Mater. Phys.*, 1996, **54**, 11169–11186.
21 G. Kresse and J. Hafner, *J. Phys.: Condens. Matter*, 1994, **6**, 8245–8257.
22 G. Kresse and D. Joubert, *Phys. Rev. B: Condens. Matter Mater. Phys.*, 1999, **59**, 1758–1775.
23 K. Momma and F. Izumi, *J. Appl. Crystallogr.*, 2011, **44**, 1272–1276.
24 X. Ren, P. Rinke, C. Joas and M. Scheffler, *J. Mater. Sci.*, 2012, **47**, 7447–7471.
25 Á. Rusznyák, V. Zólyomi, J. Kürti, S. Yang and M. Kertesz, *Phys. Rev. B: Condens. Matter Mater. Phys.*, 2005, **72**, 1–6.
26 A. Tapia, L. Aguilera, C. Cab, R. A. Medina-Esquivel, R. De Coss and G. Canto, *Carbon*, 2010, **48**, 4057–4062.

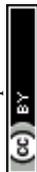